\documentclass{ws-p8-50x6-00}

\begin{document}

\title{Orbitally Excited Baryon Spectroscopy in the  $1/N_c$ Expansion}

\author{C.~L.~Schat}

\address{Jefferson Lab, 12000 Jefferson Avenue, Newport
News, VA 23606, USA \\  Department of Physics, Hampton University, Hampton, VA 23668, USA \\
E-mail: schat@jlab.org}


\maketitle

\abstracts{ The discussion of the 70-plet of negative parity baryons illustrates 
the large $N_c$ QCD approach to orbitally excited baryons. In the case of the $\ell=1$
 baryons
the existing data allows to make numerous predictions  to first order in the $SU(3)$ symmetry 
breaking. New relations between splittings 
are found that follow from the spin-flavor symmetry breaking. The $\Lambda(1405)$ is well 
described as a  three-quark state and  a spin-orbit partner of the $\Lambda(1520)$.
Singlet states with higher orbital angular momentum $\ell$ are briefly discussed. }

\section{Introduction}

In the $N_c \rightarrow \infty$ 't~Hooft limit  QCD \cite{tHo74}
 has  a contracted dynamical spin-flavor  
symmetry\cite{GeSa84} $SU(2F)_c$  for the ground state baryons\cite{Wit79} ($F$ is the number of light  flavors).
 This  is a
 consequence of unitarity in  pion-nucleon scattering in that limit and at fixed energy of 
order ${\cal O}(N_c^0)$ \cite{Dashen1,Jenk1}. In general $SU(2F)$  is broken at ${\cal O}(1/N_c)$ but
for some observables only  at ${\cal O}(1/N_c^2)$ \cite{Jenk1}, which  makes 
 the  $1/N_c$ expansion around  the  $SU(2F)$ symmetric limit
 a powerful tool of analysis as it is shown in numerous works\cite{Jenk1,CGO94,LM94,JL95,Man98}. 
The excited baryons are expected to  reveal further  the details of strong QCD and are therefore  of current 
theoretical interest\cite{SGS02,Sim02,Lee02} and also  a central goal of lattice QCD studies\cite{Ric01}.    
 In the context of the $1/N_c$ expansion\cite{CGO94,Goi97,PY98} this baryon sector is  less well understood, 
the principal reason being that even in the $N_c \rightarrow \infty$ limit the spin-flavor 
symmetry is broken\cite{Goi97}. However, most of the known baryons of negative parity
seem to fit very well in the $(3,70)$ irreducible representation (irrep) of $O(3)\otimes SU(6)$.
The $1/N_c$ operator expansion for the full 70-plet can be implemented along the lines developed for two
flavors\cite{Goi97,CCGL} and  shows that the  leading order
spin-flavor breaking (${\cal{O}}(N_c^0)$) is  indeed small, thus justifying  $SU(2F)$ as an approximate symmetry 
useful for classifying  excited baryons\cite{SGS02}.
 
 The $1/N_c$  expansion is an  appropriate tool for the study of some long-standing problems of the quark  model  in a model independent way.
 For a long time the quark model in its different versions has been the preferred framework for investigating the 
properties of baryons\cite{CandR}. Despite the
success of this model in reproducing  general features of the spectrum,   it is not a complete 
representation of QCD. One consequence of this incompleteness is that, in
those cases where the quark model does not agree with phenomenology, such as the problem of the mass splittings
between
spin-orbit partners in the negative parity baryons (spin-orbit puzzle), it is not clear whether the problem is due
to the quark model itself or to specific dynamical properties of the states involved. This situation is clarified
in the $1/N_c$ expansion where 
the presence of other operators, the leading one being  of ${\cal{O}}(N_c^0)$, solves the contradictions that arise in  
the quark model when the spin-orbit interaction is considered.   

\section{The space of states}

The  states in the  $(3, 70)$ of  $O(3)\otimes SU(6)$ decompose into five
octets ($^{2 S+1}d_J =$ $^28_{1/2}$, $^28_{3/2}$, $^48_{1/2}$, $^48_{3/2}$ and $^48_{5/2}$,  
where $S$ is the total spin, $d$ the degeneracy of the $SU(3)$
irrep and $J$ is the total angular momentum), two decuplets ($^210_{1/2}$ and $^210_{3/2}$) 
and two singlets  ($^21_{1/2}$ and $^21_{3/2}$).
An explicit representation of these  states can be obtained from a tensor product of quark states. 
Coupling an orbitally excited quark with $\ell=1$ to $N_c-1$ s-wave quarks that constitute
 a spin-flavor  symmetric core gives the following states with  core spin  $S^c$
\bea \label{wfsc}
 |J,J_z \ ; \ S&;&(\lambda,\mu), Y,I,I_z \ ; \ S^c> \ =  \nonumber \\
&=& \sum 
\left(
	\begin{array}{cc|c}
		S   & \ell & J   \\
		S_z & m & J_z
	\end{array}
\right)
\left(
	\begin{array}{cc|c}
		S^c   & \frac{1}{2} & S   \\
		S^c_z & s_z         & S_z
	\end{array}
\right)
\left(
	\begin{array}{cc|c}
		(\lambda^{c} , \mu^{c})  &  (1 , 0)                    &  (\lambda , \mu )  \\
		(Y^c, I^c, I_z^c)        &  (y ,  \frac{1}{2}  , i_z ) &  (Y, I, I_z)
	\end{array}
\right)  \nonumber \\
& & \ \ \ \ \ \ \ \ \ \ 
\left|
	\begin{array}{cc}
		S^{c}   & (\lambda^{c} , \mu^{c})  \\
		S^{c}_z & (Y^c, I^c, I_z^c)
	\end{array}
\right\rangle
\left|
	\begin{array}{cc}
		\frac{1}{2}  & (1 , 0)  \\
		s_z          & (y ,  \frac{1}{2}  , i_z )
	\end{array}
\right\rangle 
\left|
	\begin{array}{c}
		\ell   \\
		m
	\end{array}
\right\rangle \ .
\eea 
\noindent
The $(\lambda,\mu)$ labels indicate the $SU(3)$ irrep, $Y$ is the
hypercharge, $I$ the isospin and $J_z$, $I_z$ the obvious projections.
For arbitrary $N_c$ the (3,70) states are embedded in a larger multiplet and  are taken  
to have strangeness of order $N_c^0$.
From the decomposition of the $SU(6)$ symmetric
representation into irreps  of $SU(2) \otimes SU(3)$  the relations $\lambda^c + 2 \mu^c = N_c -1$ and
$\lambda^c = 2 S^c$ follow. They are the generalization of the $I=J$ rule well known for two flavors.
The (3,70) states are in the mixed symmetric irrep of $SU(6)$, which for  the  octets with $S=1/2$ 
corresponds to a linear combination  of  states of the form of Eq.(\ref{wfsc})
\bea
|^28> &=& -\frac{\sqrt{3}}{2} \sqrt{1-\frac{1}{N_c}} \ |S^c=0> \ + \ \  \frac{1}{2}\sqrt{1+\frac{3}{N_c}}\  |S^c=1> \ ,
\eea
where the  coefficients can be obtained diagonalizing the quadratic  Casimir operator of  $SU(6)$
\bea
 C^{(2)}_{SU(6)} = 2 \ G_{ha} G_{ha} + \frac{1}{2} C_{SU(3)}^{(2)} + \frac{1}{3} C_{SU(2)} \ .
\eea 
The core state for the  $^4 8$ and $^2 10$ irreps
is $|1,(2,\frac{N_c-3}{2})>$, in the $^2 1$ irrep  the core state is $
|0,(0,\frac{N_c-1}{2})>$ and the corresponding states given by Eq.(\ref{wfsc}) 
are already in the mixed representation of $SU(6)$.

The physical states are in general a mixing of states with the same $J$. In the $SU(3)$ 
symmetric limit 
only the octets with $J=1/2$ and $J=3/2$ mix. The mixing angle $\theta_{2 J }$ is defined as   
\bea
\left(
	\begin{array}{c}
		8_{J}  \\
		8'_{J}
	\end{array}
\right)
&=& 
\left(
	\begin{array}{rr}
		\cos{\theta_{2 J }}  & \sin{\theta_{2 J }}  \\
	        - \sin{\theta_{2 J }}  &  \cos{\theta_{2 J }}
	\end{array}
\right)
\left(
	\begin{array}{c}
		^28_{J}  \\
		^48_{J}
	\end{array}
\right) \ .
\eea

\section{Construction of operators}

A basis of mass operators can be built using the generators of $O(3) \otimes SU(2 F )$\cite{Goi97}.
A generic $n$-body mass operator has the general structure
\bea
O^{(n)} &=& \frac{1}{N_c^{n-1}} \ O_{\ell} \ O_q \ O_c \ ,
\eea
where the factors $O_{\ell}$, $ O_q$, and  $O_c$ can be expressed in terms of products of generators of
orbital angular momentum ($\ell_i)$,  spin-flavor of the excited quark ($s_i, t_a$ and $g_{ia} \equiv s_i t_a$) and
spin-flavor of the core ($S^c_i, T^c_a$ and $G^c_{ia} \equiv \sum_{m=1}^{N_c-1} s^{(m)}_i t^{(m)}_a$), respectively.
The explicit $1/N_c$ factors originate in the $n-1$ gluon exchanges required to
give rise to an $n$-body operator.
The matrix elements of operators may also carry a nontrivial $N_c$ dependence due to coherence
effects\cite{GeSa84,Dashen1}:
for the states considered, $G^c_{ia}$ ($a=1,2,3$) and $T^c_8$  have matrix elements of  ${\cal{O}}(N_c)$,
while  the rest of the generators have matrix elements of higher order.

In the case of the $\ell=1$  baryons the highest orbital  angular momentum operator 
that contributes is  the 
rank 2 tensor
\bea
\ell^{(2)}_{hk} &=& \frac{1}{2} \{\ell_h,\ell_k\} - \frac{\ell^2}{3} \delta_{hk} \ .
\eea

\section{Counting the number of operators}

For $N_c=3$ and in the $SU(3)$ symmetric limit there are eleven independent quantities: nine masses (one 
for each $SU(3)$ multiplet)  and two mixing
angles $\theta_1$ and $\theta_3$, which correspond to the mixing of the $^28_J$ and $^48_J$
octets  with $J=1/2$ and $J=3/2$.
This leads to the basis of eleven $SU(3)$-singlet
mass operators which are listed in Table~\ref{tab1}. Further information about the structure of these operators
can be obtained from the $SU(3)$ singlets  in the decomposition of 
\bea
\overline{70} \otimes 70 & = & 4 (1,1) \oplus 5 (1,3) \oplus 2 (1,5) \oplus (1,7) \oplus  ... 
\eea 
which shows that there are  four  operators with $\ell = 0$  ($O_1,O_6,O_7,O_{10}$), five  operators with  $\ell = 1$  
($O_2,O_4,O_5,O_9,O_{11}$), 
two  operators with   $\ell = 2$ ($O_3,O_8$) and one operator with  $\ell = 3$  that 
does not contribute in the case of interest.
In terms of $1/N_c$ one operator is of ${\cal O}(N_c)$, namely the identity,
 $O_{2,3,4}$ are  of ${\cal O}(N_c^0)$,  and the remaining seven $O_{5,...,11}$ are of ${\cal O}(1/N_c)$,
one of which is the very important hyperfine operator. They are a simple generalization of
those known for two flavors, although the calculation of their matrix elements is in general more involved. 

When $SU(3)$ breaking is included with  isospin conservation, the number of independent
observables raises up to 50, of which 30  are masses and 20  are
mixing angles.  However, if $SU(3)$ symmetry breaking is restricted to linear order in quark masses
 only isosinglet octet
operators can appear, and the number  of independent observables is reduced to 35 (21 masses and
14 mixing angles) implying  24  linearly independent octet mass operators.
As a consequence of this reduction several mass relations exist, among them there is a Gell-Mann Okubo 
relation for each
octet and an equal spacing rule for each decuplet.
The octet contributions are proportional to
$\ \epsilon~\propto~(m_s~-~m_{u,d}~)/\nu_{H}$ where $\nu_{H}$ is a typical
hadronic mass scale, for instance $m_\rho$;  for $N_c=3$ the quantity $\epsilon$ counts as of the same order as
$1/N_c$. Explicit construction shows that up to order ${\cal O}(\epsilon N_c^0)$ only a small subset
of independent octet operators $B_i$ appears. Since such octet operators are isospin singlets, it is possible to modify
them by adding singlet operators so that the resulting operators vanish in the subspace of
non-strange baryons. This procedure of improving the flavor breaking operators may change the $1/N_c$ counting: for
instance, after improving $T_8$ with the identity operator $O_1$ the resulting operator is of order $N_c^0$. Indeed, the
improved operators give the splitting due to $SU(3)$ breaking with respect to the non-strange baryons in each
multiplet, and they must be of zeroth order  or higher in $1/N_c$ for states with strangeness of order $N_c^0$.
The  four improved flavor breaking operators $\bar B_1$ through $\bar B_4$ that remain
at ${\cal O}(\epsilon N_c^0)$ when $N_c=3$ are shown in Table~\ref{tab1}.

\section{Fitting the data}

As a result of the  above analysis  the 70-plet  mass operator up to  ${\cal O}(\epsilon N_c^0)$  
has the most general form:
\\
\bea
M_{70} & =& \sum_{i=1}^{11} c_i O_i + \sum_{i=1}^{4} d_i \bar B_i~~~~,
\eea
\\
where $c_i$ and $d_i$ are unknown coefficients which are reduced matrix elements (of a QCD operator) that are 
not determined by the spin-flavor symmetry. Calculating these reduced matrix elements is
equivalent to solve QCD in this baryon sector. Fortunately, the experimental data available
in the case of the 70-plet is enough to obtain them by making a fit\cite{SGS02}. The resulting values 
are given in Table~\ref{tab1}.  The natural size of coefficients associated with the singlet operators is
set by the coefficient of $O_1$, and is about  $500 \ {\rm MeV}$, while the natural size for the
coefficients associated with octet operators is roughly $ \epsilon $ times $ 500  \ {\rm MeV}  $.
  The experimental masses (three or more stars status in the the Particle Data listing\cite{PDG}) 
 shown in Table~\ref{tab2}  together with the two leading order
mixing angles $\theta_1= 0.61$,  $\theta_3= 3.04$ \cite{HLC75,ik78} are the 19 empirical quantities used in 
the fit.  The
resulting  $\chi^2$  per degree of freedom  turns out to be $\chi^2 /4 = 1.29$. The best fit masses and 
state compositions are displayed in Table~\ref{tab2}.
\begin{table}[t]
\caption{ Operator list and best fit coefficients{\protect \cite{SGS02}}.}\label{tab1}
\begin{center}
\footnotesize
\begin{tabular}{llrrr}
\hline 
\hline
Operator & \multicolumn{4}{c}{Fitted coef. [MeV]}\\
\hline
\hline
$O_1 = N_c \ 1 $ & $c_1 =$  & 449 & $\pm$ & 2 $\ $  \\
\hline
$O_2 = l_h \ s_h$ & $c_2 =$ & 52 & $\pm$ & 15   $\ $ \\
$O_3 = \frac{3}{N_c} \ l^{(2)}_{hk} \ g_{ha} \ G^c_{ka} $ & $c_3 =$  & 116 & $\pm$ & 44  $\ $ \\
$O_4 = \frac{4}{N_c+1} \ l_h \ t_a \ G^c_{ha}$ & $c_4 =$  & 110 & $\pm$ &  16 $\ $\\
\hline
$O_5 = \frac{1}{N_c} \ l_h \ S^c_h$ & $c_5 =$  & 74 & $\pm$ & 30 $\ $\\
$O_6 = \frac{1}{N_c} \ S^c_h \ S^c_h$ & $c_6 =$  & 480 &  $\pm$ & 15 $\ $\\
$O_7 = \frac{1}{N_c} \ s_h \ S^c_h$ & $c_7 =$ & -159 &  $\pm$ & 50 $\ $ \\
$O_8 = \frac{1}{N_c} \ l^{(2)}_{hk} s_h \ S^c_k$ & $c_8  =$  & 6  & $\pm$ &   110   $\ $\\
$O_9 = \frac{1}{N_c^2} \ l_h \ g_{ka} \{ S^c_k ,  G^c_{ha} \} $ & $c_9 =$ &  213 &  $\pm$ &  153  $\ $\\
$O_{10} = \frac{1}{N_c^2} t_a \{ S^c_h ,  G^c_{ha} \}$ & $c_{10} =$  & -168 &  $\pm$ &  56  $\ $\\
$O_{11} = \frac{1}{N_c^2} \ l_h \ g_{ha} \{ S^c_k ,  G^c_{ka} \}$ & $c_{11} =$ & -133 &  $\pm$ &  130  $\ $\\
\hline
\hline
$\bar B_1 = t_8 - \frac{1}{2 \sqrt{3} N_c} O_1$ & $d_1 =$  & -81 & $\pm$ & 36 $\ $\\
$\bar B_2 = T_8^c - \frac{N_c-1}{2 \sqrt{3} N_c } O_1 $  & $ d_2 = $  & -194 & $\pm$ & 17  $\ $\\
$\bar B_3 = \frac{1}{N_c} \  d_{8ab}  \ g_{ha} \ G^c_{hb}  + \frac{N_c^2 -9}{16 \sqrt{3} N_c^2 (N_c-1)} O_1 +$ &  & & $\ $\\
\hspace*{1cm} $+ \frac{1}{4 \sqrt{3} (N_c-1)} O_6 + \frac{1}{12 \sqrt{3}} O_7 $  & $ d_3 = $  & -150 & $\pm$ & 301  $\ $\\
$\bar B_4 = l_h \ g_{h8} - \frac{1}{2 \sqrt{3}} O_2 $ & $ d_4 = $  & -82 & $\pm$ & 57  $\ $\\
\hline \hline
\end{tabular}
\end{center}
\end{table}

\section{Splitting relations}

Because at 
${\cal O}(\epsilon N_c^0)$ there are only four flavor breaking operators,
 it is possible to find  new mass splitting relations which are independent of the coefficients $d_i$. These 
relations involve states in different $SU(3)$ multiplets.
Of particular interest are the following five relations that result when the
operator $\bar B_3$ is neglected (from the fit it is apparent that $\bar B_3$ gives very small contributions):
\bea \label{splrel}
9 (s_{\Sigma_{1/2}} +  s_{\Sigma'_{1/2}}) + 21 s_{\Lambda_{5/2}} &=&
17 (s_{\Lambda_{1/2}} +  s_{\Lambda'_{1/2}}) + 5 s_{\Sigma_{5/2}} \ , \nonumber \\
2 (s_{\Lambda_{3/2}} +  s_{\Lambda'_{3/2}}) &=&
3  s_{\Lambda_{5/2}} + s_{\Sigma_{5/2}} \ , \nonumber \\
18 (s_{\Sigma_{3/2}} + s_{\Sigma'_{3/2}}) +
33 s_{\Lambda_{5/2}} &=& 28 (s_{\Lambda_{1/2}} + s_{\Lambda'_{1/2}})
+ 13 s_{\Sigma_{5/2}} \nonumber \ , \\
9 s_{\Sigma''_{1/2}} &=&  s_{\Lambda_{1/2}} +   s_{\Lambda'_{1/2}}
+ 3 s_{\Lambda_{5/2}} + 4 s_{\Sigma_{5/2}} \nonumber  \ , \\
18 s_{\Sigma''_{3/2}} + 3 s_{\Lambda_{5/2}} &=& 8(s_{\Lambda_{1/2}}
+s_{\Lambda'_{1/2}})  + 5 s_{\Sigma_{5/2}} \ .
\eea
Here $s_{{\cal B}_i}$ is the mass splitting between the baryon
${\cal B}_i$ and the non-strange baryons in the $SU(3)$ multiplet
to which it belongs. These relations are independent of mixings because they result from relations among traces of
the octet operators. If $\bar B_3$ is not neglected there are instead four relations.
The first relation in equation  (\ref{splrel})  predicts the $\Sigma_{1/2}$ to be $103 \ {\rm MeV}$ above the $N_{1/2}$, 
consistent with
the $\Sigma(1620)$,  a two star state that is not included as an input to  the fit. Each of the remaining relations makes
a similar prediction for other states but requires further experimental data to be tested.

\renewcommand{\arraystretch}{.9}
\begin{table}[t]
\caption{ Masses and spin-flavor content as predicted by the large $N_c$ analysis{\protect \cite{SGS02}}. Also given are
the empirical masses and those obtained in a quark model (QM) calculation {\protect \cite{ik78}}. }\label{tab2}
\begin{center}
\footnotesize
\begin{tabular}{ccccccccccc}\hline \hline
       & & \multicolumn{3}{c}{Masses [MeV]} & &  \multicolumn{4}{c}{Spin-flavor content} \\
State &\hspace*{.5cm} & Expt. & Large $ N_c$  &  QM &\hspace*{.5cm} & $^21$ & $^28$ & $^48$ & $^210$ \\
\hline
  $N_{1/2}$      & &$ 1538 \pm 18  $ & 1541  & 1490  &  &               &  0.82        &  0.57  &         \\
 $\Lambda_{1/2}$ & &$ 1670 \pm 10  $ & 1667  & 1650  &  & -0.21  &  0.90  &  0.37  &         \\
 $\Sigma_{1/2}$  & &$  (1620)      $ & 1637  & 1650  &  &               &  0.52  &  0.81  &  0.27   \\
 $\Xi_{1/2}$     & &$              $ & 1779  & 1780  &  &       &  0.85  &  0.44 &  0.29  \\
\hline
 $N_{3/2}  $     & &$ 1523 \pm 8   $ & 1532  & 1535  &  &       &-0.99   & 0.10   &         \\
 $\Lambda_{3/2}$ & &$ 1690 \pm 5   $ & 1676  & 1690  &  & 0.18  & -0.98  & 0.09   &         \\
 $\Sigma_{3/2}$  & &$ 1675 \pm 10  $ & 1667  & 1675  &  &       & -0.98  & -0.01   &-0.19 \\
 $\Xi_{3/2}$     & &$ 1823 \pm 5   $ & 1815  & 1800  &  &       & -0.98  &  0.03  & -0.19   \\
\hline
 $N'_{1/2}  $    & &$ 1660 \pm 20  $ & 1660  & 1655  &  &       & -0.57  &  0.82  &         \\
 $\Lambda'_{1/2}$& &$ 1785 \pm 65  $ & 1806  & 1800  &  & 0.10  & -0.38  & 0.92  &         \\
 $\Sigma'_{1/2}$ & &$ 1765 \pm 35  $ & 1755  & 1750  &  &       & -0.83  &  0.54  &  0.17   \\

 $\Xi'_{1/2}$    & &$              $ & 1927  & 1900  &  &       & -0.46  &  0.87  &  0.18   \\
\hline
 $N'_{3/2}  $    & &$ 1700 \pm 50  $ & 1699  & 1745  &  &       & -0.10  & -0.99  &         \\
 $\Lambda'_{3/2}$& &$              $ & 1864  & 1880  &  &0.01   & -0.09  & -0.99  &         \\
 $\Sigma'_{3/2}$ & &$              $ & 1769  & 1815  &  &       & 0.01   & (-0.57)  & (-0.82)   \\
 $\Xi'_{3/2}$    & &$              $ & 1980  & 1985  &  &       & -0.02  & (-0.57)  & (-0.82)   \\
\hline
 $N_{5/2}  $     & &$ 1678 \pm 8   $ & 1671  & 1670  &  &       &        &  1.00  &         \\
 $\Lambda_{5/2}$ & &$ 1820 \pm 10  $ & 1836  & 1815  &  &       &        &  1.00  &         \\
 $\Sigma_{5/2}$  & &$ 1775 \pm 5   $ & 1784  & 1760  &  &       &        &  1.00  &         \\
 $\Xi_{5/2}$     & &$              $ & 1974  & 1930  &  &       &        &  1.00  &         \\
\hline
 $\Delta_{1/2}$  & &$ 1645 \pm 30  $ & 1645  & 1685  &  &       &        &        &  1.00   \\
 $\Sigma''_{1/2}$  & &$              $ & 1784  & 1810  &  &       &  -0.14  &  -0.31  &  0.94   \\
 $\Xi''_{1/2}$     & &$              $ & 1922  & 1930  &  &       &  -0.14  &  -0.31  &  0.94   \\
 $\Omega_{1/2}$  & &$              $ & 2061  & 2020  &  &       &        &        &  1.00   \\
\hline
 $\Delta_{3/2}$  & &$ 1720 \pm 50  $ & 1720  & 1685  &  &       &        &        &  1.00   \\
 $\Sigma''_{3/2}$  & &$              $ & 1847  & 1805  &  &       &  -0.19  &  (-0.80)  &  (0.57)   \\
 $\Xi''_{3/2}$     & &$              $ & 1973  & 1920  &  &       &  -0.19  &  (-0.80)  &  (0.57)   \\
 $\Omega_{3/2}$    & &$              $ & 2100  & 2020  &  &       &             &             &  1.00   \\
\hline
 $\Lambda''_{1/2}$ & &$ 1407 \pm 4   $ & 1407  & 1490  &  & 0.97  & 0.23  & 0.04  &         \\
\hline
 $\Lambda''_{3/2}$ & &$ 1520 \pm 1   $ & 1520  & 1490  &  & 0.98  & 0.18  & -0.01  &         \\
\hline \hline
\end{tabular}
\end{center}
\end{table}

\section{The singlet Lambdas}

The  singlet Lambdas are the two lightest states of the 70-plet, something that has  
 its natural explanation  in the dominant effect of the hyperfine interaction\cite{DGG75}. 
Although spin-flavor symmetry  is broken  at  ${\cal O}(N_c^0)$, it is apparent  from our fit 
that the ${\cal O}(N_c^0)$ operators
are dynamically suppressed as their coefficients are substantially smaller than  the natural size. It turns out that
the chief contribution to spin-flavor breaking stems from  the ${\cal O}(1/N_c)$ hyperfine operator  $O_6$, as in
 the ground state baryons. Since $O_6$ is purely a core operator, the gross
spin-flavor structure of levels is determined by the two possible  core states. 
 In particular, the two singlet $\Lambda$s are not affected by $O_6$, while  the other
states are moved upwards, explaining in a transparent way the lightness of these two states. Indeed, by keeping
only $O_1$ and $O_6$ the $^2 8$  masses are $1510 \ {\rm MeV}$, the $^4 8$ and $^2 10$ masses are $1670 \ {\rm
MeV}$, and the $^2 1$ masses are left at the bottom with $1350  \ {\rm MeV}$. This clearly  shows  the dominant
pattern of spin-flavor breaking observed in the 70-plet.

The  long standing  problem in the quark model  of  reconciling the large $\Lambda(1520)-\Lambda(1405)$ splitting
with the  splittings between  the other spin-orbit partners in the 70-plet is resolved  in the large $N_c$ analysis.
The singlet $\Lambda$s  receive contributions to their masses  from $O_1$ and $\ell \cdot s$ 
while the rest of the operators 
give vanishing contributions because  the core of the singlets carries $S^c=0$. 
The splitting between the singlets is, therefore,  a clear display of the spin-orbit coupling. 
The problem with the splittings between spin-orbit partners in the non-singlet sector, illustrated  by the fact that 
the  $\ell \cdot s $ operator gives  a contribution to the $\Delta_{1/2}-\Delta_{3/2} $ splitting  that is of  
opposite sign
 of what is observed,  is now solved by the presence of the operators $O_4$, $O_5$, $O_9$ and $O_{11}$, with the  
contribution from  $O_4$ being the dominant one in accordance with the $1/N_c$ counting.
While $O_2$ and $O_4$ are of order $N_c^0$ separately, their sum  $ O_2 + O_4$ is of order $1/N_c$ 
{\em for the non-singlet states}, as can be 
seen from the explicit expressions for their matrix elements given in Table~\ref{tab3}. $O_4$ is 
therefore the natural operator  that cancels the effect of $O_2$ at large $N_c$. 
This also leaves $O_3$ as the dominant contribution to the leading mixing angles $\theta_1$, $\theta_3$. 
The 
analytic expressions for the rest of the operators will be  given elsewhere.

\begin{table}[t]
\caption{ Matrix elements of $O_2$ and $O_4$ . }\label{tab3}
\begin{center}
\footnotesize
\begin{tabular}{l}
\begin{tabular}{|c|c|c|c|}
\hline
  &  $8_{1/2}$ &  $8'_{1/2}$  &  $8_{1/2} - 8'_{1/2}$ 
\\[1.0ex]
\hline
$O_2$  & $ -\frac{2 N_c - 3}{3 N_c} $ &  $-\frac{5}{6} $ &   $-\frac{1}{3 \sqrt{2}} \sqrt{ 1 + \frac{3}{N_c}}$  \\
$O_4$  & $\frac{2}{9}  \frac{(N_c + 3)(3 N_c-2)}{N_c(N_c+1)}$ & $\frac{5}{18} \frac{3 N_c + 1}{N_c+1} $  &
          $\frac{1}{9\sqrt{2}} \frac{ 3 N_c - 5}{N_c+1}  \sqrt{1 + \frac{3}{N_c} }$ \\
\hline
\hline
   & $8_{3/2}$ &   $8'_{3/2}$ &  $8_{3/2} - 8'_{3/2}$ 
\\[1.0ex]
\hline
$O_2$ &   $\frac{2 N_c - 3}{6 N_c}$ &   $-\frac{1}{3} $     &      $-\frac{\sqrt{5}}{6}  \sqrt{1 + \frac{3}{N_c} }$ \\
$O_4$ &   $ -\frac{1}{9} \frac{(N_c+3)(3 N_c-2)}{N_c(N_c+1)} $ &   $\frac{1}{9} \frac{3 N_c + 1}{ N_c + 1}$  &  
          $\frac{\sqrt{5}}{18} \frac{ 3 N_c - 5}{N_c+1}  \sqrt{1 + \frac{3}{N_c} } $ \\
\hline
\end{tabular} \\ \\
\begin{tabular}{|c|c|c|c|c|c|}
\hline
 & $8_{5/2}$ & $10_{1/2}$ &  $10_{3/2}$  & $1_{1/2}$ 
& $1_{3/2}$  
\\[1.0ex]
\hline
$O_2$ &  $ \frac{1}{2} $  &  $\frac{1}{3}$ &           $-\frac{1}{6}$  & $ -1 $ &     $\frac{1}{2}$  \\[1.0ex]
$O_4$ &  $-\frac{1}{6} \frac{3 N_c + 1}{N_c + 1}$  &  $-\frac{1}{9} \frac{3 N_c + 7}{N_c + 1}$ & $\frac{1}{18} 
\frac{3 N_c + 7}{N_c + 1}$ & $0$ &  $0$ \\[1.0ex]
\hline
\end{tabular}
\end{tabular}
\end{center}
\end{table}

In principle, a similar situation would be expected  for states with one quark 
excited at higher angular momentum  $\ell > 1$. It is 
 interesting to note that the splittings of the observed states 
$( \Lambda (1405)\frac{1}{2}^-,\Lambda (1520)\frac{3}{2}^- )$,
$( \Lambda (1890)\frac{3}{2}^+,\Lambda (2110)\frac{5}{2}^+ )$,
$( \Lambda (1830)\frac{5}{2}^-,\Lambda (2100)\frac{7}{2}^- )$,
$( \Lambda (2020)\frac{7}{2}^+,\Lambda (2350)\frac{9}{2}^+ ) $
are in a relation $3.0:5.7:7.0:8.6$ while the $\ell \cdot s$ operator
predicts   $3.0:5.0:7.0:9.0$. Thus, the observed data also hints that $c_2$ may be of 
approximately the same size in different spin-flavor multiplets. 
Further support to this picture can be drawn from scaling down to the strange sector 
the mass splitting 
between the $(\Lambda_c(2593) \frac{1}{2}^-,\Lambda_c(2625)\frac{3}{2}^-)$ as suggested by Isgur\cite{Isg95}.

\section{Conclusions}
The $1/N_c$ expansion provides a systematic approach to the spectroscopy of
the excited  baryons. In the case of the negative parity $\ell = 1$ baryons  it 
successfully describes the existing data and, to the order considered,  also makes 
numerous testable predictions. In addition to the well known Gell-Mann-Okubo  and equal spacing  relations,
new splitting relations between different multiplets that follow from the spin-flavor symmetry  have been found.
The $\Lambda(1405)$ is well described as a
three-quark state and the spin-orbit partner of the $\Lambda(1520)$. 
Available experimental data for higher $\ell$ states and extrapolations from the charmed sector also 
seem to hint at the presence of a spin-orbit interaction. 
Effective interactions that correspond to flavor quantum number exchanges,
such as the ones mediated by the operators $O_3$ and $O_4$, are apparently needed.
Although the corresponding coefficients seem to be dynamically suppressed their relevance shows up in the well
established finer effects, namely mixings and splittings between non-singlet spin-orbit partners.
These interactions are not accounted for in the standard quark model based on one gluon exchange.

\section*{Acknowledgments} 
The results reported here were obtained in collaboration with J.L.~Goity and N.N.~Scoccola.
 I want to thank the organizers for inviting me and also to acknowledge 
 partial support from the Institute for Nuclear Theory at the University of Washington and   
from CONICET(Argentina).

\end{document}